\documentclass[aps,prb,twocolumn,superscriptaddress]{revtex4-1}
\usepackage{graphicx,amssymb,amsmath,amsfonts,verbatim,color}

\begin{document}
\bibliographystyle{apsrev}

\title{Single-crystal superconducting nanowires of NbSe$_2$ fabricated by reactive plasma etching}

\author{Shaun A. Mills}
\affiliation{Department of Physics and Materials Research Institute, The Pennsylvania State University, University Park, PA 16802, USA}

\author{Neal E. Staley}
\altaffiliation{Current Address: Department of Physics, Massachusetts Institute of Technology, Cambridge, MA, 02139}
\affiliation{Department of Physics and Materials Research Institute, The Pennsylvania State University, University Park, PA 16802, USA}

\author{Jacob J. Wisser}
\affiliation{Department of Physics and Materials Research Institute, The Pennsylvania State University, University Park, PA 16802, USA}

\author{Chenyi Shen}
\author{Zhuan Xu}

\affiliation{Department of Physics, Zhejiang University, Hangzhou 310027, China}

\author{Ying Liu}
\email{liu@phys.psu.edu}

\affiliation{Department of Physics and Materials Research Institute, The Pennsylvania State University, University Park, PA 16802, USA}

\affiliation{Department of Physics and Astronomy and Key Laboratory of Artificial Structures and Quantum Control (Ministry of Education), Shanghai Jiao Tong University, Shanghai 200240, China}

\begin{abstract}

We present the preparation and measurements of nanowires of single-crystal NbSe$_2$. These nanowires were prepared on ultrathin ($\lesssim10\text{ nm}$)  flakes of NbSe$_2$ mechanically exfoliated from a bulk single crystal using a process combining electron beam lithography and reactive plasma etching. The electrical contacts to the nanowires were prepared using Ti/Au. Our technique, which overcomes several limitations of methods developed previously for fabricating superconducting nanowires, also allows for the preparation of complex superconducting nanostructures with a desired geometry. Current-voltage characteristics of individual superconducting single-crystal nanowires with widths down to 30~nm and cross-sectional areas as low as 270 nm$^2$ were measured for the first time.

\end{abstract}

\maketitle

Nanoscale superconductors (wires, disks, loops, etc.) have long been a system of fundamental interest. Investigation of phase slips induced by either thermal activation (TAPS)\cite{Arutyunov1DSC,LAMH1,LAMH2,RogachevMagFields, RogachevNbNW, TianSnNanowire} or macroscopic quantum tunneling (MQT)\cite{GiordanoMQT,LauMQT,AltomareMQT,BezryadinMoGeMQT,TianSnNanowire}  in superconducting nanowires has deepened our understanding of phase coherence at a macroscopic length scale. Work on doubly-connected superconductors, such as ultrathin hollow cylinders of superconductors prepared on an insulating cylindrical substrate -- which were shown to exhibit both a destructive regime near half-integer flux quanta,\cite{LiuCylinder, WangAlCylinders, SternfeldAlCylinders} as predicted originally by de Gennes,\cite{deGennesTail}  and a quantum phase transition near the onset of the destructive regime\cite{WangAlCylinders} -- has shown that sample topology plays an important role in determining the properties of nanoscale superconductors. The destructive regime is the Little-Parks effect\cite{LittleParks} in the limit of an ultrasmall cylinder diameter. Furthermore, planar doubly-connected superconducting nanostructures exhibit a Little-Parks-de Gennes effect in which the destructive regime is manipulated through sample geometry by adding a side branch to a superconducting nanoloop.\cite{StaleyLPdG}

The preparation of nanowires studied previously is usually done by depositing a thin superconducting film on a suitable narrow substrate, resulting in amorphous,\cite{BezryadinMoGeMQT,LauMQT} granular,\cite{RogachevNbNW} or polycrystalline nanowires.\cite{AltomareFab,AltomareMQT} Consequently, superconducting weak links\cite{Tinkham} may form in the structure, complicating the interpretation of experimental results.\cite{DuanWL} Single-crystal superconducting nanowires can be achieved with molecular beam epitaxy,\cite{GuoPbFilm} but the rapid oxidation of such systems necessitates the addition of a protective capping layer for \emph{ex situ} transport measurements, which may lead to unintended changes in the device. Single-crystal nanowires can also be prepared by electrochemical deposition in porous media.\cite{TianNanowire} However, these wires are difficult to study once released from the porous medium, again because of oxidation. Therefore, the fabrication of ultrathin, single-crystal nanoscale superconductors is of significance. 

\begin{figure}[b!]
\includegraphics[scale=1]{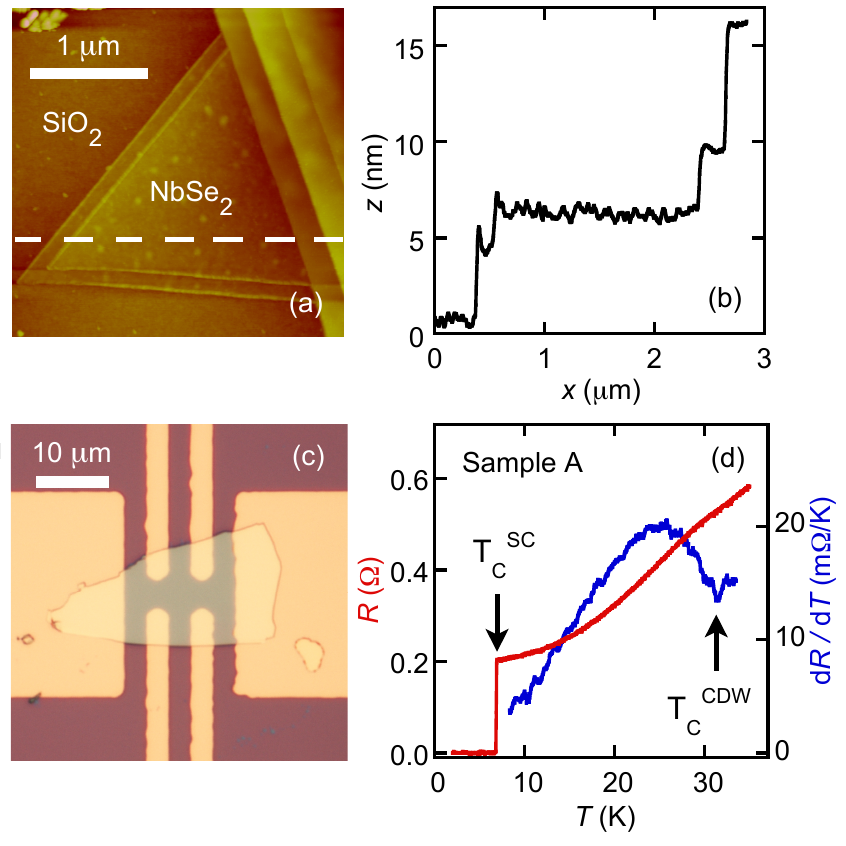}
\caption{(color online) (a) AFM image of a NbSe$_2$ flake prepared by mechanical exfoliation. The ultra-thin triangular part of the flake is connected to a thicker crystal (top-right), but devices can be fabricated on the thin regions by means of a selective etch. (b) Height profile of flake along path indicated by dashed white line in (a). (c) Optical microscopy image of NbSe$_2$ flake device prepared by photolithography. (d) Resistance (R) versus temperature (T) (red) and $dR/dT$ versus temperature (blue) measurement of device in (c). Critical temperatures for the superconducting (SC) and charge density wave (CDW) transitions are indicated.}
\label{flake}
\end{figure}

NbSe$_2$ is a layered type-II superconductor, featuring a hexagonal crystal structure with lattice constants of $a = 0.3$~nm and $c = 1.3$~nm.\cite{MarezioNbSe2Lattice} Each unit cell consists of two NbSe$_2$ layers in the AB stacking. Bulk NbSe$_2$ has an in-plane coherence length $\xi(0)=9.5$~nm and magnetic field penetration depth $\lambda=200$~nm.\cite{SanchezNbSe2Quantities} Using a mechanical exfoliation technique similar to the one developed for graphene, we isolated ultrathin flakes of NbSe$_2$ on a SiO$_2$/Si substrate. The NbSe$_2$ flakes couple to the substrate through a weak van der Waals interaction.\cite{NovoselovOrig} Flake thicknesses were determined by means of an atomic force microscope (AFM) calibrated optical color guide, which is accurate to within a single unit cell for flakes with a thickness \(\lesssim25\)~nm.\cite{StaleyNbSe2} AFM scans of typical flakes revealed extended flat terraces, often only nanometers thick, between sharp step edges (Figs. \ref{flake}(a) and \ref{flake}(b)). 

Conventional photolithography techniques were used to prepare devices of exfoliated NbSe$_2$ flakes (Fig. \ref{flake}(c)) for the measurement of their electrical transport properties (Fig. \ref{flake}(d)). All measurements presented in this article were conducted using standard DC techniques in a Physical Property Measurement System with base temperature of $1.8$~K and superconducting magnet capable of fields of $9$~T. For the sample shown in Figure \ref{flake}(c), resistance~(R) and dR/dT measurements (Fig. \ref{flake}(d)) showed a superconducting transition at $T_c=7$~K and a charge density wave (CDW) transition at $T_c^{CDW}=32$~K.\cite{InosovNbSe2CDW} The signature for the CDW transition, an abrupt change of slope in $R(T)$, indicates the high quality of our exfoliated flakes.\cite{IwayaNbSe2CDW}  Magnetoresistance measurements (data not shown) showed the upper critical field,  $H_{c_2}$, defined to be the field at which \(R(H_{c_2})=\frac{1}{2}R_N\), was $3.7$~T at $1.8$~K with the magnetic field applied along the \emph{c} axis, yielding an in-plane coherence length of $\xi(1.8\text{ K})=9.3$~nm, consistent with the bulk value. 

\begin{figure}
\includegraphics[scale=1]{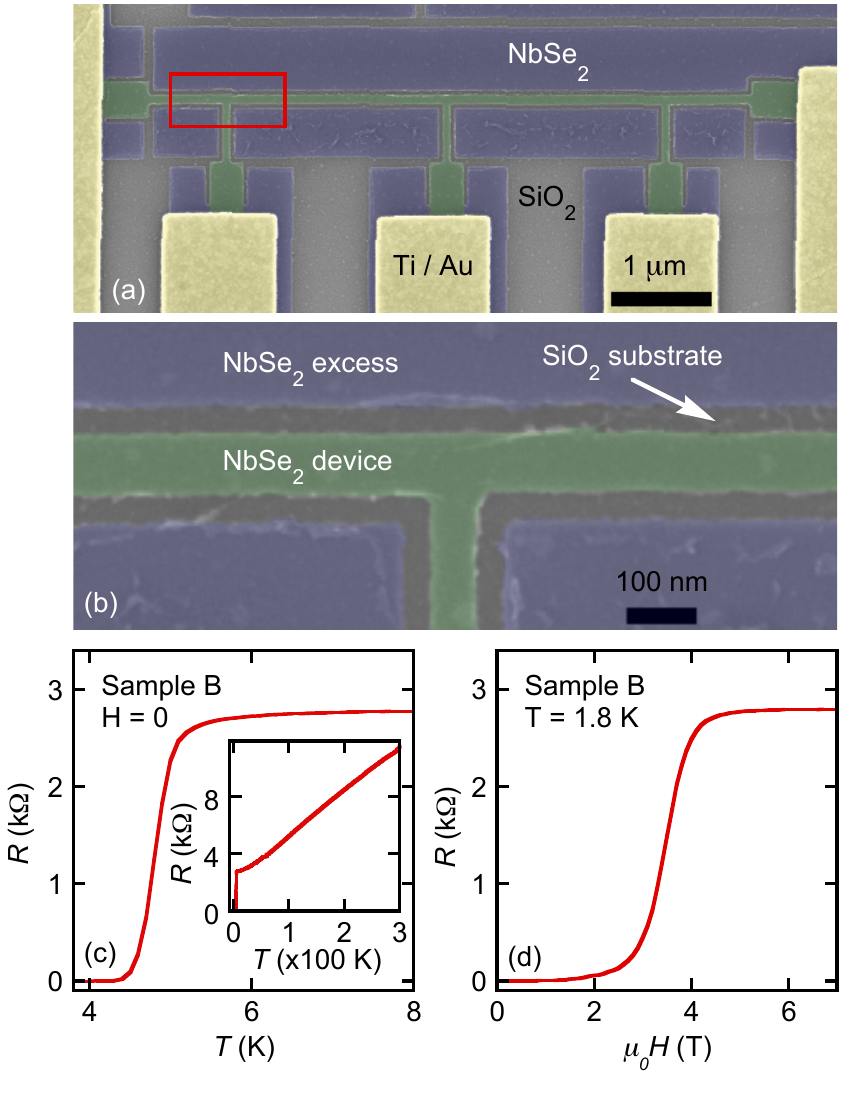}
\caption{(color online) (a) False-color scanning electron microscopy (SEM) image of a NbSe$_2$ nanowire with a width 90~nm, length 4~$\mu\mathrm{m}$, and thickness $9\pm1$~nm. (b) Expanded view of boxed region in (a). (c) Resistance (R) versus temperature (T) measurement of right half of nanowire in (a). Inset: Full range R(T). (d) Resistance versus magnetic field~(H) measurement of right half of nanowire in (a).}
\label{devices}
\end{figure}

To fabricate a nanowire, large leads for electrical transport measurements were first patterned on the exfoliated flakes using conventional electron beam lithography (EBL). The leads consisted of a $5$~nm underlayer of titanium, followed by a $30$~nm layer of gold. A second EBL step was used to outline the shape of the device. The entire flake was then subjected to a CF$_4$ reactive ion plasma etch, removing the unprotected NbSe$_2$ and leaving a structure of desired pattern. The plasma etch was performed with a parallel plate system at room temperature at a pressure of $4\times10^{-2}$~Torr with a 200~Volt DC bias and 60~Watt RF power. These parameters yielded an etch rate of $\gtrsim8$~\AA/s. An acetone bath followed by an O$_2$ plasma etch removed residual EBL resist (the CF$_4$ etch rendered a thin layer of EBL resist insoluble in acetone, necessitating the final O$_2$ etch). Scanning electron microscopy (SEM) imaging demonstrated the good quality of the devices so obtained (Figs. \ref{devices}(a) and \ref{devices}(b)). Measurements on Sample B (the rightmost 2 $\mu\mathrm{m}$ segment in Fig. \ref{devices}(a)) showed that the nanowire was superconducting with a T$_c$ of 4.8~K (Fig. \ref{devices}(c)) and critical field of 3.5~T (Fig. \ref{devices}(d)). The thicknesses of the starting flakes for Samples A and B were different (by at least a factor of 2), leading to different T$_c$ values.\cite{FrindtNbSe2UnitCell,StaleyNbSe2}  \begin{comment}\textcolor{red} {The temperature dependence of the resistance of these nanowires near T$_c$ was found to be consistent with that expected from thermally activated phase slips (TAPS), \cite{LAMH1,LAMH2} but the measured resistance deviated significantly from the TAPS theory at lower temperatures. While a combination of thermally activated and quantum phase slips has successfully explained such deviations previously, \cite{GiordanoMQT,LauMQT,AltomareMQT,BezryadinMoGeMQT,TianSnNanowire} we found that our data fitting along this direction would require using somewhat unphysical parameters.  }\end{comment}

The selective etch process allows for the fabrication of nanostructures of NbSe$_2$ of an arbitrary geometry, including doubly- and multiply-connected samples. Thus, this method has a distinct advantage over existing methods such as templating.\cite{BezryadinMoGeMQT} We were able to fabricate independent nanowires from one single-crystal flake, as seen in Figure \ref{devices}(a), thereby ensuring identical processing conditions among devices. 

\begin{figure}[t]
\includegraphics[scale=1]{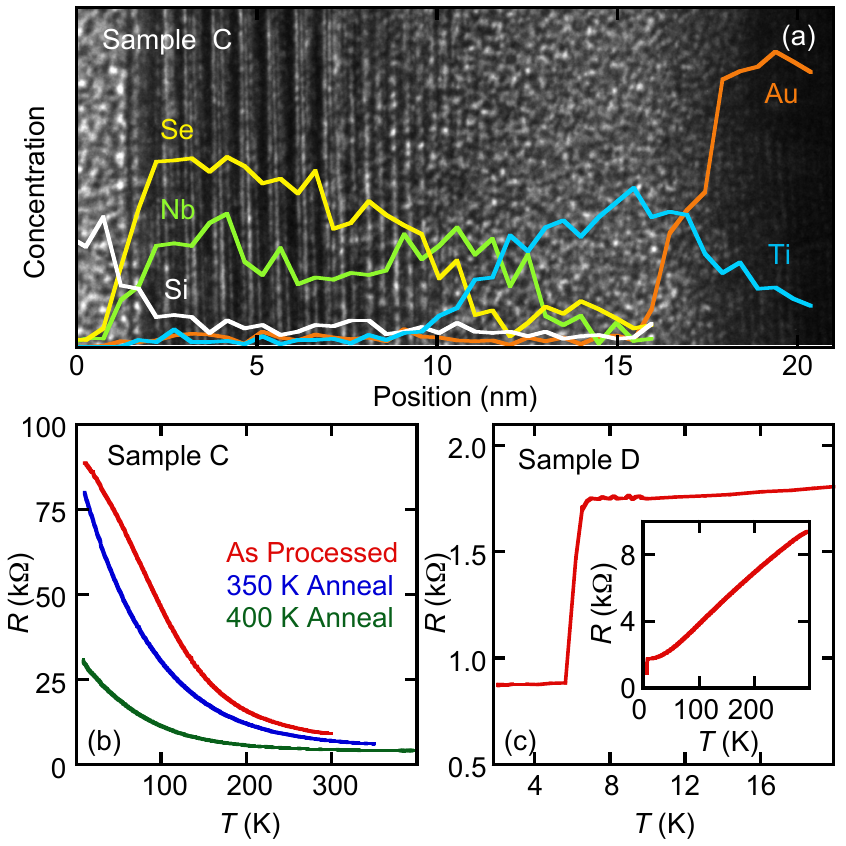}
\caption{(color online) (a) High-resolution TEM image of the contact region of Sample C after the 400~K anneal, showing (left to right) silicon substrate, pristine NbSe$_2$ flake, restructured NbSe$_2$ flake, titanium, and gold. Elemental concentrations measured by EDX with the microscope in STEM mode (5 \AA~resolution). (b) Two-terminal R(T) traces for NbSe$_2$ Sample C. Top curve (red) is as processed, middle (blue) is after a one-hour high vacuum anneal at 350~K, bottom (green) is after a similar anneal at 400~K. (c) Three-terminal resistance (R) vs. temperature (T) of a NbSe$_2$ nanowire, Sample D. Base temperature contact resistance is $\sim900$~$\Omega$. Inset: Full range R(T).}
\label{contact}
\end{figure}

For electrical transport measurements of nanowires,  low-resistance contacts are crucial. The contact issue has been the main obstacle for the electrical transport measurements of individual single crystalline nanowires of almost all elemental superconductors.\cite{WangRuContact, TianSnNanowire} In the present case, despite a carefully controlled cleaning process prior to depositing the electrical contacts, low-resistance contacts were difficult to obtain reliably. To gain insight into the behavior of the contact, we performed high resolution cross-sectional transmission electron microscopy (TEM) studies of the interface between the NbSe$_2$ flake and the Ti/Au measurement leads. TEM samples were thinned with a focused beam of gallium ions to allow the transmission of electrons. Due to the destructive nature of the TEM sample preparation, all scans were performed post-measurement. Figure~\ref{contact}(a) shows a TEM image of one contact region of Sample C. The superimposed lines indicate the relative concentrations of elements measured using energy-dispersive x-ray spectroscopy (EDX) with the TEM in scanning mode. Interestingly, Ti was found to diffuse into both the NbSe$_2$ flake and Au overlay. Furthermore, the crystal lattice of the flake abruptly restructures within about 5~nm of the top surface, and the top-most layers of the flake show a pronounced Se deficiency.  While the cause of the restructuring of the NbSe$_2$ top layers is not known, it appears to be unrelated to Ti inter diffusion, as the same restructuring was found in portions of flake not covered by Ti. It is possible the restructuring results from a high-vacuum anneal this device was subjected to (discussed below), especially considering the lack of restructuring seen in similar, yet non-annealed, samples. The Se-poor surface region may result from the 180$^\circ$C bake all samples were subjected to in order to cure the EBL resist. Interestingly, recent work on ultra-thin exfoliated NbSe$_2$ flakes suggests flakes thinner than 4 unit cells are non-conducting after EBL processing.\cite{ElBanaFlakes} Prior work utilizing a lithography-free technique, however, observed superconductivity in flakes down to a single unit cell.\cite{StaleyNbSe2} It is possible the brief EBL bake causes the Se deficiency and negatively impacts the electrical properties of the topmost layers of material.

\begin{figure}[b!]
\includegraphics[scale=1]{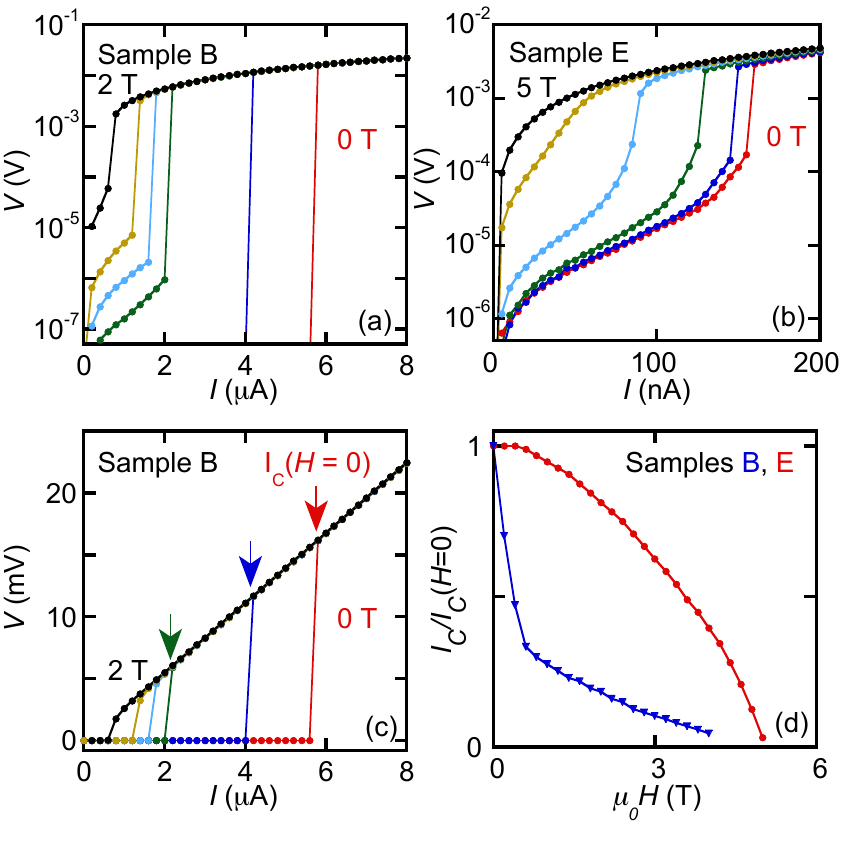}
\caption{(color online) (a) Voltage (V) versus current (I) for Sample B in an applied magnetic field of (right to left) 0, 0.2, 0.4, 0.6, 1.0, and 2.0 T ($H\parallel c$) at 1.8~K. Solid lines are guides to the eye. (b) V(I) for Sample E in an applied magnetic field of (right to left) 0, 1, 2, 3, 4, and 5 T ($H\parallel c$) at 1.8~K. Solid lines are guides to the eye. (c) Same as (a) but on a linear scale. Arrows indicate critical current, $I_c$, at 0, 0.2, and 0.4~T. (d) Normalized $I_c$ versus applied field ($\mu_0H$) for Samples B (blue triangles) and E (red circles) with $H\parallel c$ at 1.8~K. Solid lines are guides to the eye.}
\label{characterization}
\end{figure}

Vacuum annealing does seem to improve the contact to our devices. In Figure \ref{contact}(b), the two-terminal resistance of Sample C reveals insulating behavior over the entire temperature range and a base temperature contact resistance of nearly 100~k$\Omega$. We annealed the device \emph{in situ} under high vacuum (\(\mathrm{P}\lesssim10^{-5}\)~Torr), first at 350~K, then at 400~K. The resulting R(T) traces are also shown in Figure \ref{contact}(b). While annealing does reduce the contact resistance, it may lead to crystal restructuring as pointed out above. Fortunately, as shown in Figure~\ref{contact}(c), low contact resistance can be achieved even without annealing the sample. Here, the three-terminal resistance of another NbSe$_2$ nanowire measured at low temperatures is plotted. After the superconducting drop at $\sim7$~K, the residual contact resistance is roughly 900 $\Omega$, which is among the lowest we have measured. 

In Figure \ref{characterization}, we show four-terminal current-voltage characteristics with a magnetic field applied along the $c$ axis for two nanowires: Sample B of width 90~nm and Sample E of width 30~nm. Both nanowires are 2~$\mu$m long and 9~nm thick. At low fields, Sample B switches discontinuously between a superconducting state of unmeasurably low resistance and a fully normal resistive state (Fig.~\ref{characterization}(a)). At higher fields, where we expect stronger superconducting fluctuations, a finite resistance tail persists at low bias currents. This tail first appears at a field on the same order of magnitude as the one at which Abrikosov vortices are stable in a nanowire of this width.\cite{KoganLondon} However, a comparison with Sample E (Fig. 4(b)) suggests the finite resistance state is not strictly due to Abrikosov vortices -- Sample E should be too narrow to sustain vortices, and, furthermore, it exhibits a low bias resistive state even in zero external field. While the resistive tail is also not described satisfactorily by the theory of MQT or TAPS, it may be attributable to other superconducting fluctuations,\cite{LarkinSCFluctuations} which should be more pronounced in the more geometrically constrained nanowire.

In the high voltage regime, the nanowires exhibit a sharp departure from normal state resistance at a well defined current, $I_c$ (see Fig.~\ref{characterization}(c)). The evolution of $I_c$ as a function of applied field is shown in Figure~\ref{characterization}(d). The kink in the $I_c(H)$ trace for Sample~B may be related to vortex entry,\cite{BenkraoudaIc} especially considering the lack of a corresponding kink for the narrower Sample~E. This kink is also absent when the magnetic field is applied in the $ab$ plane (data not shown). In this orientation, the ratio $w/\xi$ is comparable to the ratio for Sample E, making vortex entry again unlikely.

The fabrication process we have presented allows for the study of superconductivity in single crystals of arbitrary geometries. While we have achieved low-resistance contacts, the exact mechanism responsible is not well understood. We have utilized this process to investigate nanowires of single-crystal NbSe$_2$, observing a kink in the evolution of the critical current with magnetic field for samples wide enough to allow vortex entry.

The nano fabrication part of this work is supported by the National Science Foundation (NSF) under Grant DMR 0908700 and Penn State MRI Nanofabrication Lab under NSF Cooperative Agreement 0335765, NNIN with Cornell University. The measurement part is carried out using support from the Department of Energy under Grant No. DE-FG02-04ER46159. Y.L. also acknowledges support from NSFC (Grant 11274229) and MOST of China (Grant 2012CB927403).

\end{document}